# Virtualization Technology: Cross-VM Cache Side Channel Attacks make it Vulnerable

**Abid Shahzad**
School of Engineering, Computer and Mathematical Sciences
Auckland University of Technology
Auckland, New Zealand
Email: ashahzad@aut.ac.nz

**Alan Litchfield**
Service and Cloud Computing Research Lab
Auckland University of Technology
Auckland, New Zealand
Email: alitchfi@aut.ac.nz

## Abstract

Cloud computing provides an effective business model for the deployment of IT infrastructure, platform, and software services. Often, facilities are outsourced to cloud providers and this offers the service consumer virtualization technologies without the added cost burden of development. However, virtualization introduces serious threats to service delivery such as Denial of Service (DoS) attacks, Cross-VM Cache Side Channel attacks, Hypervisor Escape and Hyper-jacking. One of the most sophisticated forms of attack is the cross-VM cache side channel attack that exploits shared cache memory between VMs. A cache side channel attack results in side channel data leakage, such as cryptographic keys. Various techniques used by the attackers to launch cache side channel attack are presented, as is a critical analysis of countermeasures against cache side channel attacks.

## 1 Introduction

In this paper we address specific concerns for the management of secure computing in a Cloud Computing (CC) environment. The paper presents the results of a systematic review of factors relating to cross-VM cache side channel attacks and countermeasures to prevent attacks or mitigate successful intrusions.

CC is marketed as Infrastructure-as-a-Service (IaaS), Platform-as-a-Service (PaaS) and Software-as-a-Service (SaaS) to customers (Modi et al. 2013; Hashizume et al. 2013; Chhabra and Dixit 2015; You et al. 2012; Srinivasan et al. 2012; Ayala et al. 2013). It offers individuals and organizations the use of outsourced services through the Internet and typically through a pay-per-use model, as opposed to buying and setting up resources in-house (Khorshed et al 2012; Modi et al. 2013; Chhabra and Dixit 2015; Srinivasan et al. 2012; Vaquero et al. 2011).

Besides providing a cost effective solution, the CC environment also gives benefits such as elasticity (Khorshed et al 2012; You et al. 2012; Srinivasan et al. 2012; Ayala et al. 2013; Tianfield 2012), scalability (Hashizume et al. 2013; Chhabra and Dixit 2015; Vaquero et al. 2011) and multi-tenancy (Hashizume et al. 2013; Chhabra and Dixit 2015; Tianfield 2012).

Virtualization is core to CC (Shoaib and Das 2014) and helps Cloud Service Providers (CSPs) create a multi-tenanted architecture. Virtualization is the simulation of hardware and/or software upon which different other software and applications can run. Virtualization allows the sharing of hardware resources between different tenants by creating Virtual Machines (VM) (Chhabra and Dixit 2015) that emulate the physical server system. For example, by using virtualization in IaaS, customers can create, share, copy and migrate VMs (Hashizume et al. 2013) and install different operating systems, and install and run software according to their unique requirements (Chhabra and Dixit 2015). The virtualization management system, Virtual Machine Monitor (VMM) ensures that the VMs running on the shared hardware are isolated from each other as separate entities (You et al. 2012).

A VMM or hypervisor is the software layer between hardware and the host operating system of the hardware server (You et al. 2012). The VMM sets the platform for virtualization by allowing the creation and management of different VMs on a server. The hypervisor controls the flow of instructions between the guest VM operating system and the physical hardware (Ayala et al. 2013), involving elements such as CPU core, cache memory, main memory, hard drives and network interface cards.

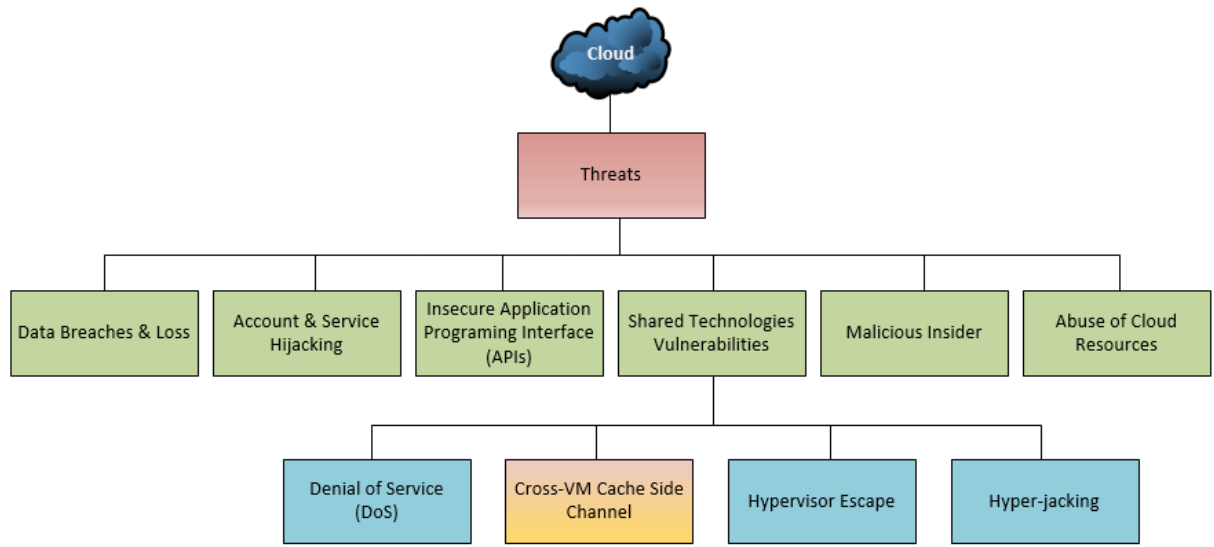

*Figure 1: Threats and Vulnerabilities to CC*

Hypervisors have two broad categories: a) Native or bare-metal virtualization hypervisors, such as XenServer, Microsoft Hyper-V and VMware ESX/EXSi; b) Hosted virtualization hypervisor, for example VMware Workstation/Player, Kernel-based VM (KVM), Microsoft Virtual PC and Oracle VirtualBox (Pek et al. 2013). However, a detailed analysis by Diego et al. (2013) shows that vulnerabilities exist in the popular open source XEN and KVM hypervisors. These have been reported in vulnerability databases; National Vulnerability Database (NVD), Red Hat Bugzilla, Security Focus and CSV Details. The vulnerabilities present in these hypervisors are characterized and furthermore, the categories of vulnerabilities are integrated to identify attack paths. This provides the opportunity for countermeasures to be developed.

The rest of the paper is organized as follows. A discussion about known threats and vulnerabilities in each cloud service model is presented in section 2. Section 3 explains the cross-VM side-channel attack. In section 4, cache side channel attack is explained along with, its different types and the techniques used to launch these attacks. We discuss cache side-channel attacks in the virtualized environment in section 5. Section 6 presents the results of an analysis of existing countermeasures proposed to defend against these attacks.

## 2    Threats and Vulnerabilities in Cloud Computing

In this section, we discuss threats and vulnerabilities in CC, some common threats and vulnerabilities are highlighted in detail, and account and service hijacking.

*Vulnerability* refers to a weakness or flaw in a system that can be exploited by an event called an *attack* (Hashizume et al. 2013; Tianfield 2012). A *threat* (see Figure 1) is an exploitation of any known vulnerability that can result in serious loss of data and information (Modi et al. 2013). These threats and vulnerabilities are perceived as a barrier for organizations wishing to adopt cloud services (Hashizume et al. 2013). The Cloud Security Alliance (CSA) has highlighted top threats to CC (Alva et al. 2013).

*Data Breaches and Loss* are very severe threats because of cloud services' shared architecture. Data loss can occur if data are being deleted without having backup or if there is any disaster and no proper mechanisms are adopted to recover it (Khorshed et al 2012).

Confidentiality, integrity and availability of customer's data are put at risk (Alva et al. 2013) when an attacker has a chance to hijack a user's account details by stealing user credentials and information. An attacker may use various methods such as phishing, DoS and Man-in-the-Middle attack to steal user credentials (Khorshed et al. 2012) and then use them to eavesdrop on a user's activities and transactions. More sophisticated attacks involve combinations of these methods for pecuniary gain. Some methods for which successful attacks may be achieved include:

- *Insecure Application Programing Interfaces (APIs)*: A CSP offers infrastructure, software and platform services to customers and gives them access to services through Application Programing Interfaces (APIs) (Khorshed et al 2012; Chhabra and Dixit 2015; Ayala et al. 2013). Users manage services they consume by using these APIs (Modi et al. 2013), for example, PaaS APIs provide access and functionality for cloud, SaaS APIs provide connectivity to applications, IaaS APIs provides control and distribution over specific cloud resources.

- Confidentiality, integrity and authentication of a cloud service are very much dependent on how APIs are designed and developed (Hashizume et al. 2013; Vaquero et al. 2011; Ayala et al. 2013). Thus, it is the responsibility of the designer and developer to make it difficult for malicious users to exploit potential vulnerabilities. For example, design flaws of the APIs to the underlying SSL libraries (Georgiev et al. 2012) provide leverage for a user with malicious intent to execute harmful actions (Srinivasan et al. 2012).

- *Malicious Insider:* Confidentiality, integrity and availability may be violated if customer data abuse occurs from someone with authorized access but with malicious intent (Alva et al. 2013). The malicious insider threat may become apparent because of a lack of transparency in CSP processes and procedures and the level of access each employee has to customer data (Modi et al. 2013; Srinivasan et al. 2012).

- *Abuse of Cloud Resources:* CSPs give free trials of IaaS to customers in order to convince them to purchase this service later on. However, due to a lack of control exercised by the CSP, customers may take the opportunity to perform malicious activities using the powerful hardware platform provided by the CSP (Khorshed et al. 2012; Modi et al. 2013). Furthermore, such hardware gives them chance to crack encryption keys (Modi et al. 2013; Zhang et al. 2013).

- *Shared Technology Vulnerabilities:* Virtualization provides multi-tenancy through the sharing of hardware resources like CPU cores, high level cache, storage devices and network interface cards among different tenants (Ayala et al. 2013). The hypervisor provides a platform for virtualization and is responsible for isolation between different VMs running on a shared server.

By exploiting vulnerabilities or configuration flaws, an attacker can gain unauthorized access to a hypervisor to control a cloud platform and exploit other customers' VMs (Khorshed et al. 2012; Modi et al. 2013). For example, infrastructure sharing is one core cloud service but because of hardware sharing limitations, lacks basic protection mechanisms to protect customer network traffic, data and other applications. This provides an attacker a chance to hijack user credentials, eavesdrop on information and control other users' VMs (Khorshed et al. 2012).

Shah and Vania (2014) highlight two common vulnerabilities present in virtualization technology, VM hopping (co-resident attack) and VM Migration/Mobility vulnerabilities. The VM live migration process exposes guest OS information on the fly. Through VM hopping an attacker can extract information from a co-resident VM by exploiting the shared cache memory among different VMs that are running co-resident with the attacker's VM.

## 3   Cross-VM Side Channel Attacks

Cross-VM side channel attacks are identified as a very sophisticated attack in CC. In cases where there exist shared hardware resources, the side channel attack exploits information obtained from the usage of, for example, Central Processing Unit (CPU) core and high level cache memory as opposed to exploiting theoretical weaknesses like the brute force attack (Vaquero et al. 2011).

In a typical cross-VM side channel attack scenario, an adversary places a malicious VM co-resident to the target VM so that they share the same hardware resources (Ristenpart et al. 2009). Then, the attacker extracts useful information such as cryptographic keys from the target VM (Zhang et al. 2012) to use them for traffic eavesdropping and man-in-the-middle attacks. Through the side channel attack, an attacker sharing the same cache as the victim can monitor the cache access behaviour of the victim. For example, the attacker is able to monitor cache timing information by measuring the execution of different operations on the victim's VM (Zhang and Lee 2014). Generally, the attacker exploits timings in the shared high-level cache memory. However, power consumption or electromagnetic leaks can also be used as a vector to launch side channel attacks (Vaquero et al. 2011).

## 4　Cache-Based Side Channel Attacks

### 4.1　Cache Architecture

The modern x64 microprocessor has multiple levels of cache: Level 1, Level 2 and Level 3. In a processor with multiple cores and three levels of cache, each processor core has its own separate L1 and L2 cache while L3 cache is shared among all the cores of the processor (Yarom and Falkner 2013; Kim et al. 2012; Zhang et al. 2011). Therefore, the shared architecture of L3 cache among different cores creates an opportunity for a malicious user to exploit it and launch a side channel attack.

In a typical scenario, when the processor needs data, it checks L1, L2 and L3 caches respectively. If the requested data are present at any level then the data are read and a cache hit occurs. However, if data are not available in any cache level, then a cache miss occurs and the processor gets data from RAM and places it in the cache ready for the next call. In the event of a cache miss, the processor may also take advantage of spatial locality by fetching data from near cache locations, because it is expected that data in near cache locations are going to be accessed soon (Irazoqui et al., 2014a; 2014b). Moreover, data fetched from RAM and placed in cache are in fixed sized blocks called *cache lines,* and a cache entry is created during this process.

### 4.2　Cache Organization

To gain an understanding of cache organisation and the role of cache-hypervisor relationship, types of cache organisation are described.

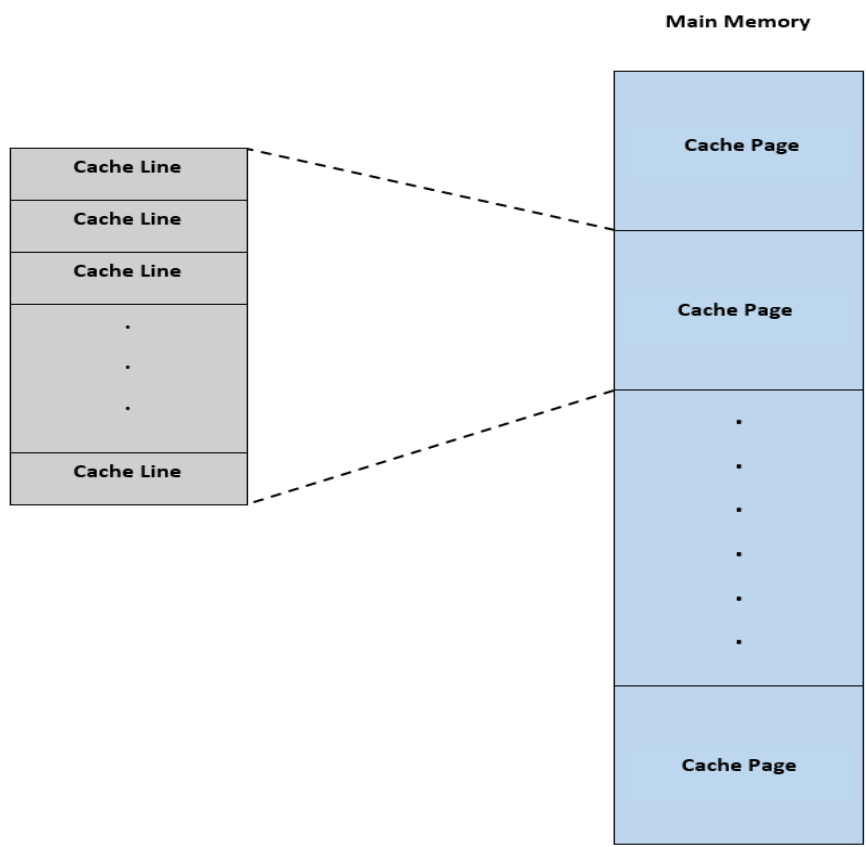

*Figure 2: Cache Organisation*

Cache consists of cache pages and cache lines (see Figure 2 on the previous page). The main memory (RAM) is divided into a number of cache pages. The size of each cache page depends on the overall size of cache. The cache page is further divided into a number of fixed size cache lines. The type of processor and the cache design defines the size of each cache line (Intel 2014).

### 4.2.1 Fully Associative

This scheme allows only lines, not cache pages, to move from main memory to cache. Figure 3 illustrates that any line from main memory can be located at any location on the cache.

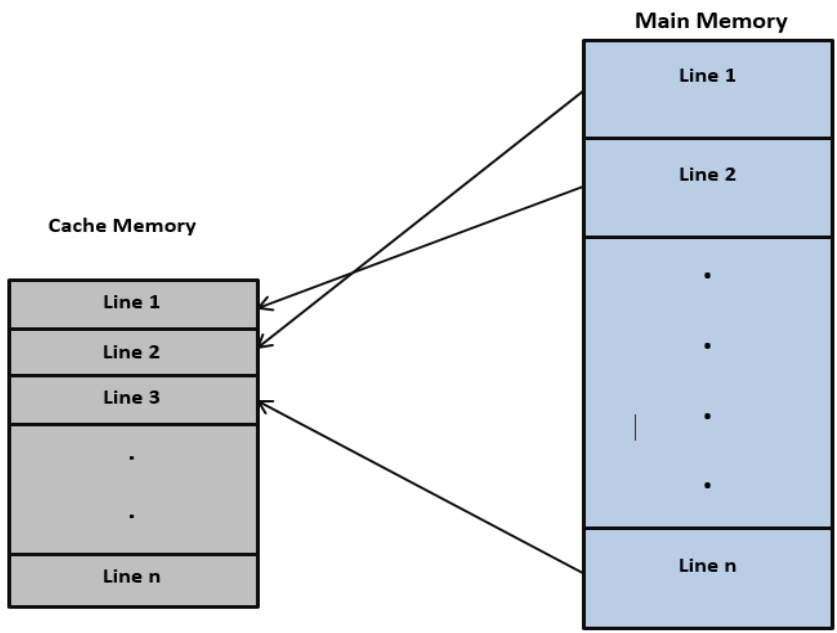

*Figure 3: Fully Associative Cache*

### 4.2.2 Direct-Mapping

Direct mapping in cache is also called 1-Way set associative cache. In this scheme, the size of each main memory page is equal to the size of the cache. Specific lines from main memory move to the same line of the cache as shown in Figure 4. For example, the first line of the first main memory page is stored in the first cache line. Later, if the second main memory page's first line needs to be stored in the first cache line, then the first page's line needs to be evicted.

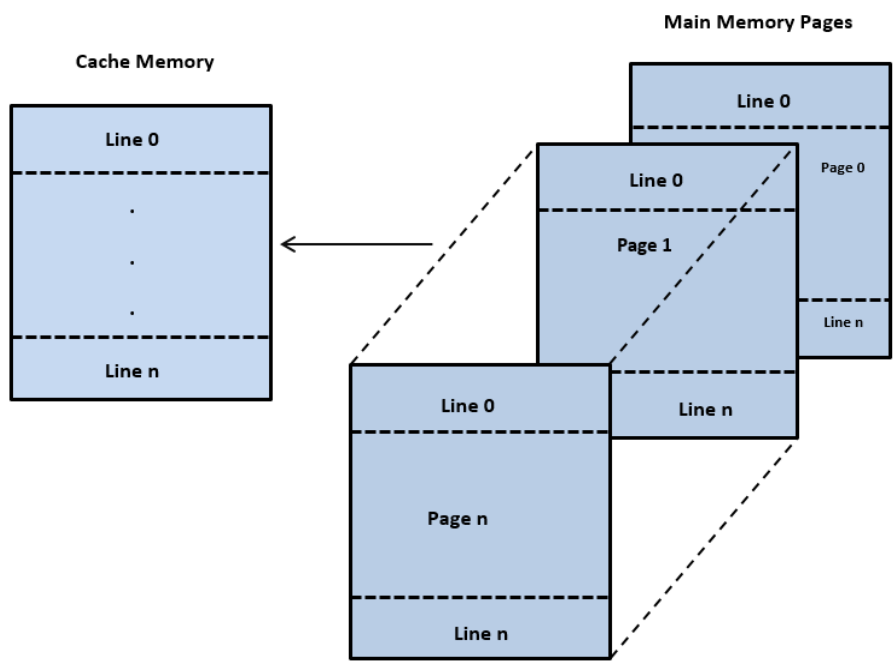

*Figure 4: Direct Mapping Cache*

#### 4.2.3 Set-Associative

Set-Associative (Figure 5) combines the above schemes and divides the cache into equal sections called ways. The size of the cache page is equal to the size of a cache way. This scheme allows 2 or 4 (typically) lines of main memory pages to be stored on the same cache lines of 2 or 4 cache ways at any time.

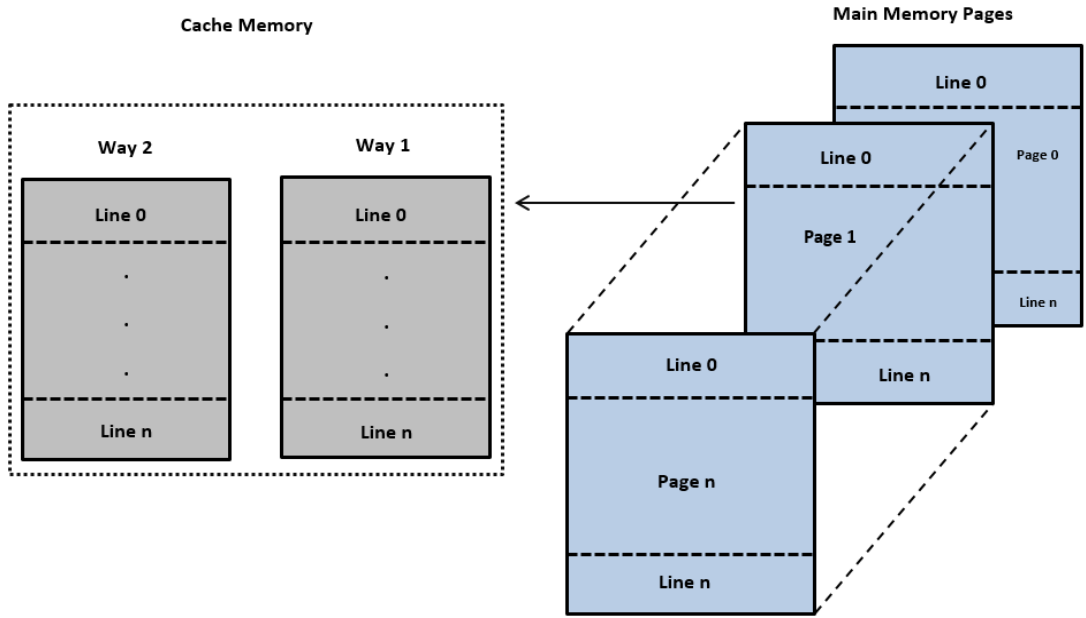

*Figure 5: Set Associative Cache*

Cache is normally configured as a set associative map (Zhang and Reiter 2013). A $w$-way set associative cache is partitioned into $m$ sets and with $w$ lines in each set. Each block of data from RAM is placed into one cache set $m$, but can be placed at any of $w$ lines within the set (Zhang et al. 2011). For data that have already been moved to cache, then if data are accessed by a cryptographic algorithm such as Advanced Encryption Standard (AES) from one of the cache lines then the cache will be accessed much faster. The algorithm's data encryption/decryption time depends on the accessed table positions and timing information, which can be exploited to gain access to the private keys used (Irazoqui et al., 2014a; 2014b). Therefore, cache timing information is important for side channel attack.

### 4.3 Cache Side Channel Attack Techniques

Side channel attacks are very sophisticated attacks on cryptographic algorithms like AES, resulting in the leakage of encryption keys used by these algorithms. In a side channel attack, an adversary exploits the information gained from hardware implementation such as shared cache memory timing, electromagnetic radiation or power consumption of the system.

Over the last few years, cache side channel attacks have been demonstrated using different techniques. For cache side channels, there are two types of technique:

- *Prime + Probe Technique*: Prime + Probe is a common technique used by an attacker in order to learn shared cache behaviour and more specifically, which cache set is accessed by the target. Explained in next section, Tromer et al. (2010), Ristenpart et al. (2009), Zhang et al. (2012), and Liu et al. (2015) use the Prime + Probe technique to launch access-driven cache side channel attacks. In order to observe cache usage by a victim's process, the attacker runs a spy process in parallel to observe the status.

- *Flush + Reload Technique*: In a Flush + Reload cache side channel attack, an attacker runs a spy process that monitors shared memory pages accessed by the victim's process. Using a spy process, the attacker can monitor the memory lines that a victim's process has evicted from the cache. By monitoring processes, the attacker is able to determine which memory lines the victim accessed. From this information, the attacker can extract information from the data processed by the victim. Yarom and Falkner (2013), Yarom and Benger (2014), and Irazoqui (2014b) use the Flush + Reload technique to demonstrate a cache side channel attack in a virtualized environment.

### 4.4 Types of Cache Side Channel Attacks

Cache based side channel attacks have three types: time-driven, access-driven and trace-driven. In all the cases, the attacker uses the shared cache to monitor the activities of another VM in a hypervisor. The types are differentiated by the means by which monitoring occurs.

#### 4.4.1 Time-driven

In this type of cache side channel attack, an attacker exploits the correlation between the cryptographic operation and cache misses of a victim (Kong et al. 2009). That is, an adversary measures the time a victim's process takes to complete a whole operation such as the time it takes to undertake a cryptographic process while memory is being accessed. This is achievable because the amount of time it takes for memory to be accessed is dependent upon the state of the cache. The attacker compares the different execution times of processes against inputs and looks for specific patterns. The difference in timings can be used as leverage to extract information about encryption keys.

#### 4.4.2 Access-driven

This attack provides an adversary with a platform to execute a malicious process in parallel with the victim's cryptographic process to derive an insight into the victim's cache behaviour (Kong et al. 2009). The attacker learns what cache sets have been accessed by the victim's cryptographic operation by evicting the victim's cache memory pages. This forces the victim to have a cache miss and then the attacker may observe cache miss behaviour with the knowledge that the operation being performed is that instigated by the victim.

#### 4.4.3 Trace-driven

In this cache side channel attack an attacker monitors the cache lines accessed in a cache set by the victim and obtains a profile of cache activity during an encryption process. This sets the platform for an attacker to measure memory lines accessed by the cryptographic operation and that resulted in a cache hit. Zhao and Wang (2010) present a trace-driven attack on AES by monitoring cache miss information and S-box misalignment; around 200 samples of cache miss trace information were enough to extract a complete 128-bit AES within seconds.

## 5 Cache Side Channel Attacks in the Virtualized Environment

In this section we provide evidence of the effectiveness of side channel attacks through the application of a number of techniques. The cache-based side channel attack presents a serious threat to the cloud virtual environment. Whereas the client server based network environment makes it difficult to undertake this type of attack, it is the shared hardware resources that characterise a typical hypervisor environment that permits the cache side channel attack. As described in the previous section, an attacker sharing the same cache as a victim can monitor the cache access behaviour of the victim, such as cache timing information, by measuring the execution of different operations on the victim's VM (Zhang and Lee 2014). Cache timing information gives an adversary a platform from which to launch a side channel attack to break cryptographic algorithms (Godfrey and Zulkernine 2013) and thus to extract encryption keys. In the virtual environment, prior to the cross-VM side channel attack, the attacker needs to identify the target VM's location and place a malicious VM co-resident with the target. Later, that attacker may use the maliciously placed VM to extract information from the target VM with the side channel attack.

Using Amazon EC2 as case study, Ristenpart et al. (2009) explains that it is possible to map the cloud infrastructure to find out that where a particular target VM resides. Two VMs are co-located on the shared hardware by exploiting EC2's placement policy and since the cache is shared, the extraction of keystroke timing information using access-driven side channel attack is enabled.

In a different example in which both access and timing approaches are used, Zhang et al. (2012) present an access-driven cross VM side channel attack on a symmetric multiprocessing system and extract useful information from a target VM. Through a cache timing attack, they recover an ElGamal decryption key from the victim VM using the libgcrypt cryptographic library.

Yarom and Falkner (2013) present a cross CPU-cores side channel attack by exploiting and monitoring memory page sharing between different processes. This attack is based on the new Flush + Reload technique that evict a memory line from the cache, allowing the attacker and target system to execute at same time. Yarom and Benger (2014) extend this and present a Flush + Reload cache side channel attack that exploits an x86 processor architecture as it allows one process to monitor another process' access to cache memory pages. The attack highlights a serious threat to any cryptographic algorithm based on OpenSSL implementation.

Using Bernstein's correlation attack as a base for the first time in a virtual environment, Irazoqui et al. (2014a) show that cross-VM side channel attacks on VMware, Xen and KVM are possible. Despite running on different CPU cores on the same hardware, VMs are vulnerable to side channel attacks resulting in the leakage of AES keys by using the timing difference between cache line accesses. Weiss et al. (2012) extend the Bernstein correlation attack by demonstrating a time-driven cache attack on an embedded ARM based operating system platform, bypassing the isolation characteristics of virtualization.

Irazoqui (2014b) use the Flush + Reload cache side channel attack on VMs running on a VMware hypervisor. The attack exploits a memory de-duplication mechanism called transparent page sharing. The result is a very fast full recovery of keys from the AES implementation of OpenSSL 1.0.1.

## 6  Cache Side Channel Attack Countermeasures

The shared cache between different cores of modern processors is a concern because it permits side channel attacks. Previously, we present an overview of side channel attacks exploiting cache memory. The problem in cache design is cache contention, which allows different users' processes to evict the contents of a cache that is used by other users (Jin et al. 2009). In this section and keeping shared cache vulnerabilities in mind, to defend against cache side channel attacks, a range of mitigation and prevention techniques or countermeasures are presented.

In Table 1, cache side channel attack countermeasures are critically analysed and present the following types of solution:

- *Hardware-based Solutions*: Page (2005), Domnitser et al. (2012), and Kong et al. (2013) propose new designs for a shared cache wherein cache portions are created and by isolating these portions from each other, each CPU core is each assigned a separate portion.

- *Software-based Solutions*: Coppens et al. (2009), Jin et al. (2009), Aviram et al. (2010), Shi et al. (2011), Zhang & Reiter (2013), and Godfrey and Zulkernine (2013) propose a software based solution to protect against side channel attacks in a client-server based network as well as a cloud environment.

- *Hypervisor-based Solutions*: Raj et al. (2009), Vattikonda et al. (2011), Kim et al. (2012), and Wu et al. (2014) propose a solution involving changes at the hypervisor level.

| Paper Title | Author | Solution Proposed | Limitations |
| --- | --- | --- | --- |
| Partitioned Cache Architecture as a Side-Channel Defence Mechanism | Page, D. | Cache partition based solution. Design makes cache the visible part of the architecture and divides it into regions. Each individual cache region is allocated to a separate application, reducing the chances of interference as compared to single shared cache | Solution requires changes to underlying hardware design that can create significant performance degradation. Moreover, the solution conflicts with the cloud model either by restricting hardware requirements or customers of CSP need to have additional skills. |

*Table 1. Analysis of Cache Side Channel Attack Countermeasures*

| Paper Title | Author | Solution Proposed | Limitations |
| --- | --- | --- | --- |
| Non-monopolizable caches: Low-complexity mitigation of cache side channel attacks | Domnitser et al. | Cache partitioning technique with minor changes in cache replacement logic without influencing overall cache core circuitry. Dynamically reserves cache lines for legitimate users by restricting an attacker to only accessing unreserved cache lines | The solution is only effective when the attacker's malicious process is running in parallel with a victim's process. Furthermore, it is difficult to find any general hardware based solutions that are adopted in mainstream processors as such solutions are tested qualitatively. |
| Architecting against software cache-based side-channel attacks | Kong et al. | Mitigation technique that hides or obfuscates cache access by different cores | The solution leads to different challenges in design modification. Requires code changes to make the solution understand branch prediction analysis based side channel attack and to protect the information. Moreover, modifications to mitigate side channel attack can impact the performance of the system |
| Practical mitigations for timing-based side-channel attacks on modern x86 processors | Coppens et al. | Compiler based side channel mitigation technique to defend against timing side channel attacks in non-cloud environment. They modify the compiler to remove dependencies of control flow on cryptographic functions, such as secret keys | The hardware solution lacks generality and gives poor performance. Furthermore proposed solution leaves cloud applications vulnerable to trace-driven attacks |
| A simple cache partitioning approach in a virtualized environment | Jin et al. | Static page colouring cache partitioning technique that is implemented in Xen VMM environment. The proposed approach restricts L2 shared cache exploitation and its implementation in VMM is fully transparent to guest VMs | The proposed approach allocates static cache pages to VMs that results in a limited number of cache sets available for use by VMs requiring cache access. This results in degradation of system performance not only for the VMs that have already secured a cache portion but also for the other VMs that need to have cache access |

*Table 1: Continued*

| Paper Title | Author | Solution Proposed | Limitations |
| --- | --- | --- | --- |
| Determinating timing channels in compute clouds | Aviram et al. | A software method to prevent timing channel based attack on a CSP enforced deterministic execution as compared to hardware based solution such as cache partitioning | The proposed solution forces the VMs execution to be deterministic to protect against timing channel attacks, resulting in many applications being left vulnerable. Needs fine grain timing information for execution in cloud environment |
| Limiting cache-based side-channel in multi-tenant cloud using dynamic page colouring | Shi et al. | Dynamic cache memory page colouring technique that provides strong cache isolation between different processes of security critical operations | From the client end, the approach requires modification to the components of a hypervisor. Prohibits the use of some features that a VM can have. Therefore, such a solution limits its adoption in a cloud environment |
| A server-side solution to cache-based side-channel attacks in the cloud | Godfrey, M. and Zulkernine, M. | A mitigation technique for a cloud server by disabling the overlapping characteristics of cache memory, ensuring isolation in the cache portion accessed by VMs. Technique uses cache flushing between prime and trigger functions | The solution limits the utilization of the available resources the VMs that results in greater costs |
| Duppel: Retrofitting commodity operating systems to mitigate cache side channels in the cloud | Zhang, Y. and Reiter, M. K. | A system called Duppel that adds noise to complicate cache timing signals. Duppel adds noise by cleaning L1 cache frequently in parallel with the execution of its tenant workload, at a speed that it adjusts by assuming what timings reflect the workload execution being monitored by another VM | In a Simultaneous Multithreading (SMT) architecture, this scheme needs the user to identify the process that needs to be protected, making the solution infeasible for protection against side channel attacks. Furthermore, it decreases cache efficiency because of the time needed to do the flush |

*Table 1: Continued*

| Paper Title | Author | Solution Proposed | Limitations |
| --- | --- | --- | --- |
| Resource management for isolation enhanced cloud services | Raj et al. | The shared cache is partitioned into several regions, with specific cache lines using page colouring. It provides isolation between different VMs by allowing them to access only certain lines by partitioning memory pages | Solution provides static cache partitioning for each VM separately which increases overhead on the system when CSP wants to run more VMs on the same physical machine |
| Eliminating fine grained timers in Xen | Vattikonda et al. | A mitigation approach by modifying RDTSC instruction on Xen based VMs. It reduces the VMs' access to cache timing information by providing a fuzzy timer instead of fine-grained timers. | The proposed solution protects against timing channel attacks. However, it is not feasible for applications running in the cloud that need fine-grained timing information |
| STEALTHMEM: System-Level Protection Against Cache-Based Side Channel Attacks in the Cloud | Kim et al. | A dynamically last level cache partitioning mechanism that protects the cache by locking cache lines per processor core. VMs running on each core loads data into these locked cache lines called stealth pages, which cannot be evicted by another VM | The solution partitions the cache and assigns a separate portion of cache to each VM. However, it needs user interaction for client side modification if some applications need access to hidden pages of cache memory. Therefore, the solution is not appropriate for the cloud |
| C2detector: a covert channel detection framework in CC | Wu et al. | Presented as a framework by making small modifications to a cloud system. It uses a captor located in the Xen hypervisor and a synthesis analyser implemented as Markov and Bayesian for detection of three types of covert channels between VMs running on same hardware server | The proposed solution requires additional hardware resources because of its complex design. It is used for policing purposes |

*Table 1: Continued*

## 7   Implications of the Proposed Solutions

The shared cache memory partitioning or isolation and assigning of a separate portion of cache memory to each VM can reduce or eliminate successful side channel attacks in a virtualized environment. Cloud service providers seek fast or instantaneous delivery of services to customers but to mitigate the effects of side channel attacks, the proposed solutions of cache partitioning require changes and modifications to the underlying hardware design. The result of changes, as they are proposed now, is an overall degradation of performance in the virtual environment. Although some solutions may apply when hardware manufacturers make design changes. The alternative to making changes to the hardware configuration is to provide a new cache partition technique may prove to be very effective.

## 8  Conclusion

Acquired from CSPs, virtualization is the core component of CC that provides isolation between the different hardware and software services of customers. However along with its benefits, virtualization has vulnerabilities. A common vulnerability in the virtualization environment is the share cache memory that permits exploitation through a Cross-VM cache based side channel attack, resulting in the leakage of AES cryptographic keys. In this paper, we present an overview of the potential for the launching of a side channel attack in a virtual environment. We also present the results of a critical analysis of solutions proposed by authors. From this analysis, we fail to see that there is an effective countermeasure to the Cross-VM cache-based side channel attack despite that there is demand for a strong defence mechanism by cloud services consumers.